\documentclass[a4paper]{jpconf}

\usepackage{epstopdf}
\usepackage{graphicx}

\def\sumint{\sum\!\!\!\!\!\!\!\int\,}
\def\Mp{M_{n,{\bf p}}}

\begin{document}

\title{On an exactly solvable confining quark model and its thermodynamics}

\author{Bruno W. Mintz\footnote{In collaboration with Let\'\i cia F. Palhares and Marcelo S. Guimar\~aes.}}

\address{Departamento de F\'\i sica Te\'orica, Instituto de F\'\i sica, UERJ 
- Universidade do Estado do Rio de Janeiro,
Rua S\~ao Francisco Xavier 524,  20550-013,  Maracan\~a,  Rio de
Janeiro,  Brasil}

\ead{brunomintz@gmail.com}

\begin{abstract}
We perform an exact computation of the grand partition function of a model of confined quarks
at arbitrary temperatures and quark chemical potentials. The model is inspired by a version of 
QCD where the perturbative BRST symmetry is broken in the infrared, while perturbative QCD is 
recovered in the ultraviolet. The theory leads, even at tree level, to a quark mass function 
compatible with nonperturbative analyses 
of lattice simulations and Dyson-Schwinger equations. In spite of being defined at tree 
level, the model produces a non-trivial and stable thermodynamic behaviour at arbitrary 
values of temperature or chemical potential. Results for the pressure and the trace 
anomaly as a function of temperature qualitatively resemble those of 
non-perturbative interactions as observed in lattice simulations. The cold and dense 
thermodynamics is also contains non-trivial features, being unlike 
a gas of free massive particles.
\end{abstract}

\section{Introduction}

Given the difficulty to address the problem of confinement in strongly interacting systems 
directly from its fundamental theory, Quantum Chromodynamics (QCD), several alternative 
approaches have been proposed. One of the main such approaches is that of effective models 
of QCD, which are quantum field theories that possess one or more fundamental aspects of 
the original theory but being nevertheless easier to have some information extracted. 
  
Regarding the quark sector of QCD, two quite sucessful models in the description of chiral 
symmetry breaking and its restoration at high temperature are the Linear Sigma Model with 
quarks (LSM) \cite{GellMann:1960np}, and the Nambu-Jona-Lasinio (NJL) Model 
\cite{Nambu:1961tp}. In their 
original formulations, these models do not address the issue of quark confinement. Indeed, 
in both models quarks are effectively treated as on-shell quasiparticles. Two possible 
directions that can be taken towards a simultaneous description of both chiral and confinement 
dynamics are represented either by the coupling of the Polyakov loop to quark degrees of 
freedom (the so-called PLSM and PNJL models) 
\cite{Meisinger:1995ih,Fukushima:2003fw,Megias:2004hj,Ratti:2005jh,Schaefer:2007pw,Schaefer:2009ui}, 
or by considering nonlocal interactions between quarks as a result of their nonperturbative coupling 
to gluons, as in the nonlocal versions of the NJL model 
\cite{Bowler:1994ir,General:2000zx,Contrera:2010kz,Benic:2012ec,Benic:2013eqa}.
Following  \cite{Baulieu:2008fy,Baulieu:2009xr,Dudal:2013vha,Capri:2014bsa,Guimaraes:2015vra}, 
we consider a third possibility, which is inspired by the (refined) Gribov-Zwanziger 
effective theory for infrared QCD \cite{Gribov:1977wm,Zwanziger:1989mf,Dudal:2008sp}, although not 
equivalent to it. 

Although it has been long clear that quarks and gluons are confined to hadrons, a definite theoretical 
criterion for confinement is not yet a settled issue. A sufficient condition for the absence of 
isolated quarks or gluons from asymptotic states (i.e., confinement) is assumption that they 
violate reflection positivity \cite{Osterwalder:1973dx}. We therefore take violation of 
reflection positivity as our criterion for confinement, following, e.g., \cite{General:2000zx,Alkofer:2000wg}. 
In this work, we explore the thermodynamics of a quark model in which confinement is encoded 
in the positivity violation of the quark propagator.

\section{A nontrivial solvable quark model}

As evidence from lattice QCD  \cite{Parappilly:2005ei} and Dyson-Schwinger Equations \cite{Aguilar:2008xm} 
studies shows, the zero-temperature quark propagator can be quite well parametrized by a 
momentum-dependent mass function compatible with the functional form
\begin{equation}\label{eq:mass-function}
 M_{eff}(p) = \frac{\Lambda}{p^2+m^2}+m_0,
\end{equation}
where $m_0$ is the quark current mass. Indeed, the data of  \cite{Parappilly:2005ei} can be 
well fitted by Eq. (\ref{eq:mass-function}), with the values $\Lambda=0.196\,$GeV$^3$, 
$m^2=0.639\,$GeV$^2$, and $m_0=0.014$MeV \cite{Dudal:2013vha}. It is interesting to notice 
that, with such parameters, the resulting euclidean quark propagator 
\begin{equation}
 S(p)=\frac{1}{\gamma\cdot p + M_{eff}(p)}
\end{equation}
displays violation of reflection positivity, indicating the confinement of quarks \cite{Capri:2014bsa}. 

The mass function (\ref{eq:mass-function}) can be obtained from the lowest-level quark propagator 
of the theory given by sum of the QCD lagrangian,
\begin{equation}
 S_{QCD} = \int d^4x\left[\frac14 F_{\mu\nu}^aF_{\mu\nu}^a + \bar\psi_\alpha^i[i(\gamma_\mu)_{\alpha\beta}D_\mu^{ij} - m_0\delta_{\alpha\beta}\delta^{ij}]\psi_\beta^j 
                + ib^a\partial_\mu A_\mu^a + \bar c^a \partial_\mu D_\mu^{ab}c^b\right],  
\end{equation}
with the BRST invariant action
\begin{eqnarray}\label{eq:auxiliary-fields}
    S_{\xi\lambda} &=& s\int d^4x\left[ -\bar\eta^i_\alpha\partial^2\xi_\alpha^i + \bar\xi^i_\alpha\partial^2\eta_\alpha^i + m^2(\bar\eta^i_\alpha\xi_\alpha^i - \bar\xi_\alpha^i\eta_\alpha^i) \right]\nonumber\\
    &=& \int d^4x\left[-\bar\lambda_\alpha^i\partial^2\xi_\alpha^i - \bar\xi^i_\alpha\partial^2\lambda_\alpha^i - \bar\eta_\alpha^i\partial^2\theta_\alpha^i + \bar\theta_\alpha^i\partial^2\eta_\alpha^i + m^2\left(\bar\lambda^i_\alpha\xi_\alpha^i + \bar\xi_\alpha^i\lambda_\alpha^i + \bar\eta_\alpha^i\theta_\alpha^i - \bar\theta_\alpha^i\eta_\alpha^i\right) \right]
\end{eqnarray}
and the coupling term
\begin{equation}\label{eq:BRST-breaking}
    S_M = \int d^4x\left[M_1^2(\bar\xi^i_\alpha\psi_\alpha^i + \bar\psi_\alpha^i\xi_\alpha^i) - M_2(\bar\lambda_\alpha^i\psi_\alpha^i + \bar\psi_\alpha^i\lambda^i_\alpha)\right]
\end{equation}
between the quark fields $\psi$ and the auxiliary fields $\xi$ and $\lambda$. 
The resulting action 
\begin{equation}\label{eq:model}
 S = S_{QCD} + S_{\xi\lambda} + S_M
\end{equation}
has been shown to be renormalizable \cite{Baulieu:2009xr}, being equivalent to 
QCD in the high-energy limit but radically changing the infrared sector of the 
theory (with respect to the perturbative picture). We interpret the extra fields 
and their interaction with the quark fields $\psi$ as a local way (in the sense 
of QFT) to take into account an effective dressing of 
quarks by gluons. The auxiliary fields of the quark sector of (\ref{eq:model}), 
can be straightforwardly integrated out, providing an effective theory for the quarks, 
whose (nonlocal) action reads, in the quadratic approximation,
\begin{equation}\label{eq:non-local-action}
 S_{nl}=\int d^4x\,\bar\psi_\alpha^i\left[i(\gamma_\mu)_{\alpha\beta}\delta^{ij}\partial_\mu - 
           \delta^{ij}\delta_{\alpha\beta}\left(\frac{2M_1^2M_2}{-\partial^2+m^2} + m_0 \right)\right]\psi^j_\beta.
\end{equation}

Notice that, in this model, the quark mass function (\ref{eq:mass-function}) is directly 
derived from the effective action (\ref{eq:non-local-action}), with $\Lambda\equiv 2M_1M_2$.

Although for now we only shall investigate the model in the quadratic level, it is perfectly possible 
to consider a loop expansion of the free energy \cite{KapustaGale}. It is also reasuring that 
the full local model (\ref{eq:model}) is a renormalizable QFT, as shown in \cite{Baulieu:2009xr}.


\section{The partition function}

At lowest order, the theory defined by (\ref{eq:model}) has a quadratic action. Therefore, 
its grand partition function can be exactly calculated using standard techniques of 
Finite-Temperature Field Theory \cite{KapustaGale}. Besides having a quadratic action, 
the theory does not correspond to a free theory, a fact that can be clearly seen, e.g., 
from the nonlocal form of the equivalent effective action (\ref{eq:non-local-action}). 

One can easily introduce temperature by compactifying one euclidean direction (conventionally 
the 4-direction), whereas the introduction of the chemical potential is more subtle. 
Starting from the local action
(\ref{eq:model}), one must first calculate the hamiltonian ${\cal H}$. By identifying the 
quark number with the charge associated with the $U(1)$ global symmetry transformation
\begin{equation}
 \psi(x)\;\;\longrightarrow\;\;e^{-i\alpha}\psi(x),
\end{equation}
one can use Noether's theorem to calculate the corresponding conserved 
current density $j^\mu$ and let ${\cal N}=\int d^3x j^0$ be the quark number operator.
The resulting grand partition function 
\begin{equation}
 Z(T,\mu) = \Tr\exp\left[-\frac{{\cal H}-\mu {\cal N}}{T}\right],
\end{equation}
can be cast in a functional integral form and straightforwardly calculated \cite{Guimaraes:2015vra}.
The result is convinently split into a $\mu-$independent term plus a $(T,\mu)-$dependent term as 
\begin{eqnarray}\label{eq:logZ_T0_and_mu}
 \frac{\log Z(T,\mu)}{2\beta VN_cN_f} &=& \sumint\log\left\{\beta^2\left[{\bf p}^2 + \Mp^2(0) +\omega_n^2\right]\right\} + \sumint\log\left\{\frac{{\bf p}^2 + \Mp^2(\mu) - (i\omega_n+\mu)^2}{{\bf p}^2 + \Mp^2(0) +\omega_n^2}\right\}\crcr
                            &\equiv&  \frac{\log Z(T,0)}{2\beta VN_cN_f} +  \frac{\log Z^{(\mu)}(T,\mu)}{2\beta VN_cN_f},
\end{eqnarray}
where we used the standard sum-integral notation,
\begin{equation}
 \sumint (\cdots) \equiv T \sum_{n=-\infty}^\infty\int\frac{d^3p}{(2\pi)^3}(\cdots).
\end{equation}

The $\mu=0$ term can be split in a sum of four terms, two of which corresponding to positive pressures of 
particles with complex conjugate masses, one of which of a particle with real mass, and one of which 
with a negative contribution or the pressure. Each of this term can be calculated straightforwardly as 
the pressure of a free gas \cite{KapustaGale}. After the usual vacuum energy subtraction \cite{KapustaGale}, 
one finds 
\begin{equation}\label{eq:log-Z_zero_mu-final}
 \log Z(T,0) = \log Z_0 + 4N_cN_f V\int\frac{d^3p}{(2\pi)^3}
      \log\left[\frac{\left(1+e^{-\beta\varphi_1}\right)\left(1+e^{-\beta\varphi_2}\right)\left(1+e^{-\beta\varphi_3}\right)}{\left(1+e^{-\beta\varphi_0}\right)^2}\right],
\end{equation}
for the $\mu=0$ contribution, where
\begin{equation}
 \log Z_0 = 2N_cN_f \beta V\int\frac{d^3p}{(2\pi)^3}\left(\varphi_1+\varphi_2+\varphi_3-2\varphi_0\right) 
\end{equation}
is the pure vacuum contribution. The quantities $\varphi_i$ ($i=0,1,2,3$) are known functions of the internal momentum $p^2$ 
\cite{Guimaraes:2015vra}. 

The $\mu-$dependent term is given by
\begin{equation}\label{eq:logZ_finite-mu}
 \log Z^{(\mu)}(T,\mu) = 2\beta VN_cN_f\sumint\log\left\{\frac{{\bf p}^2 + M_{n,{\bf p}}^2(\mu) - (i\omega_n+\mu)^2}{{\bf p}^2 + M_{n,{\bf p}}^2(0) +\omega_n^2}\right\}.
\end{equation}
A closed expression for (\ref{eq:logZ_finite-mu}) can be found at the zero-temperature limit. Using Cauchy theorem, one 
may write
\begin{eqnarray}\label{eq:logZ_finite-mu_zeroT}
  \log Z(0,\mu) &=& \log Z^{(\mu)}(0,\mu) = 2\beta VN_cN_f\int\frac{d^3p}{(2\pi)^3}\int_{-\infty}^{\infty}\frac{d\theta}{2\pi}f(i\theta+\mu)\crcr
                &=& 2\beta VN_cN_f\int\frac{d^3p}{(2\pi)^3}\int_{0}^{\infty}\frac{d\theta}{2\pi}\left[f(i\theta+\mu)+f(-i\theta+\mu)\right],
\end{eqnarray}
where
\begin{equation}
 f(\xi) := \log\left\{\frac{\Omega_{{\bf p}}^2(\xi^2) - \xi^2}{\Omega_{{\bf p}}^2[(\xi-\mu)^2] - (\xi-\mu)^2}\right\}
\end{equation}
and
\begin{equation}\label{eq:nonlocal_dispersion_relation}
\Omega_{{\bf p}}^2(\zeta):={\bf p}^2 + \left[\frac{M_3}{-\zeta+{\bf p}^2 + m^2} + m_0 \right]^2.
\end{equation}

The expressions above allowed us to evaluate the partition function (\ref{eq:logZ_T0_and_mu}) exactly in order 
to compute several thermodynamical quantities. We present our results in the next section.

\section{Results}

From the partition function, one can derive any equilibrium thermodynamical quantities. Let us first show our 
results for the pressure
\begin{equation}
 P(T,\mu) = \frac{T}{V}\log\,Z(T,\mu).
\end{equation}

\begin{figure}[h]
\begin{minipage}{17pc}
\includegraphics[width=17pc]{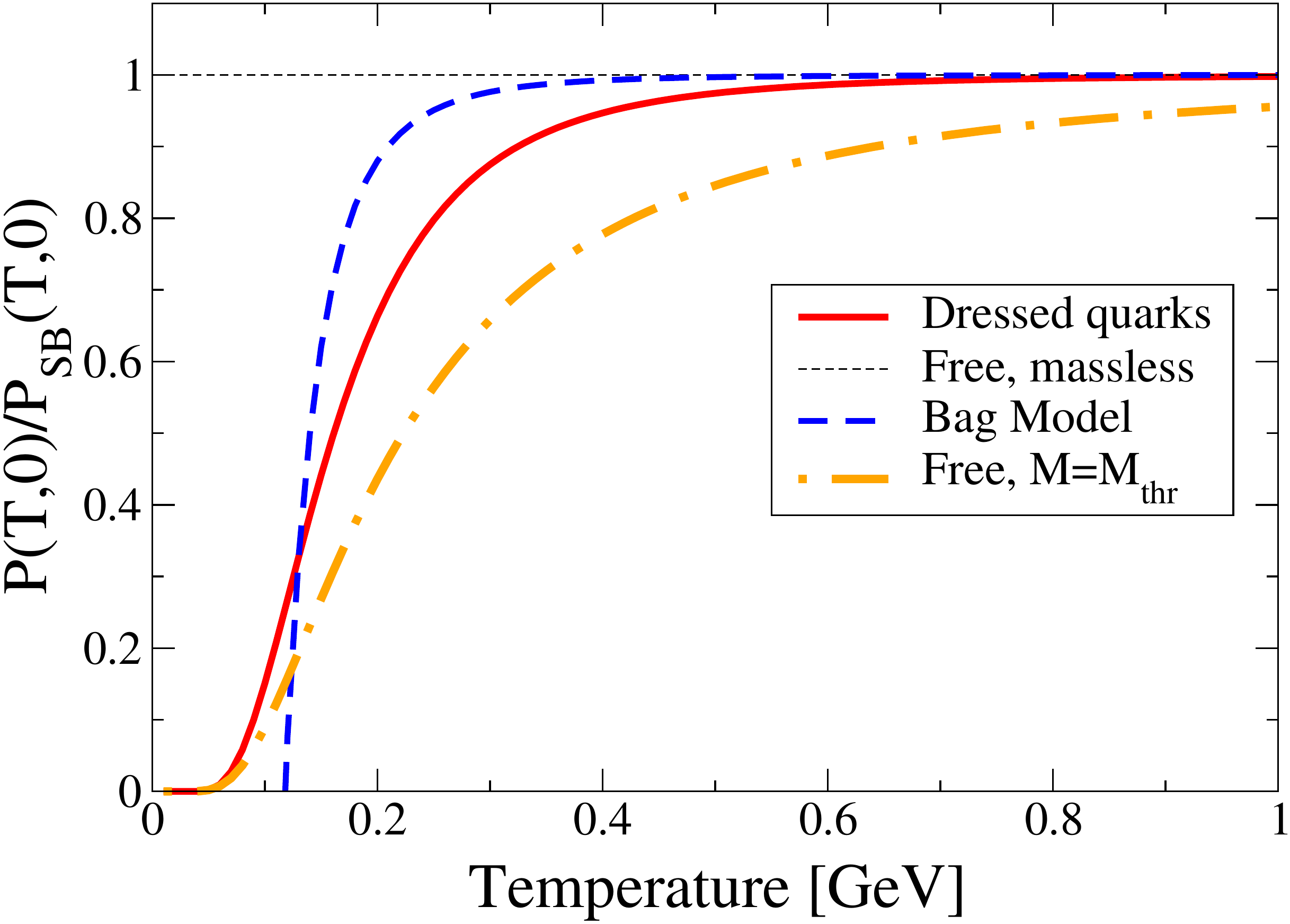}
\caption{\label{fig1} Pressure for the model quarks (solid, red line), for bag model quarks 
(dashed, blue line, with $B=(0.145\,{\rm GeV})^4$), and for a gas of free but massive quarks (dashed-dotted 
yellow line, with mass $M_{thr}=0.467$ GeV). All lines are normalized by the Stefan-Boltzmann limit 
$P_{SB}=7\pi^2N_cN_fT^4/180$.}
\end{minipage}\hspace{2pc}%
\begin{minipage}{17pc}
\includegraphics[width=17pc]{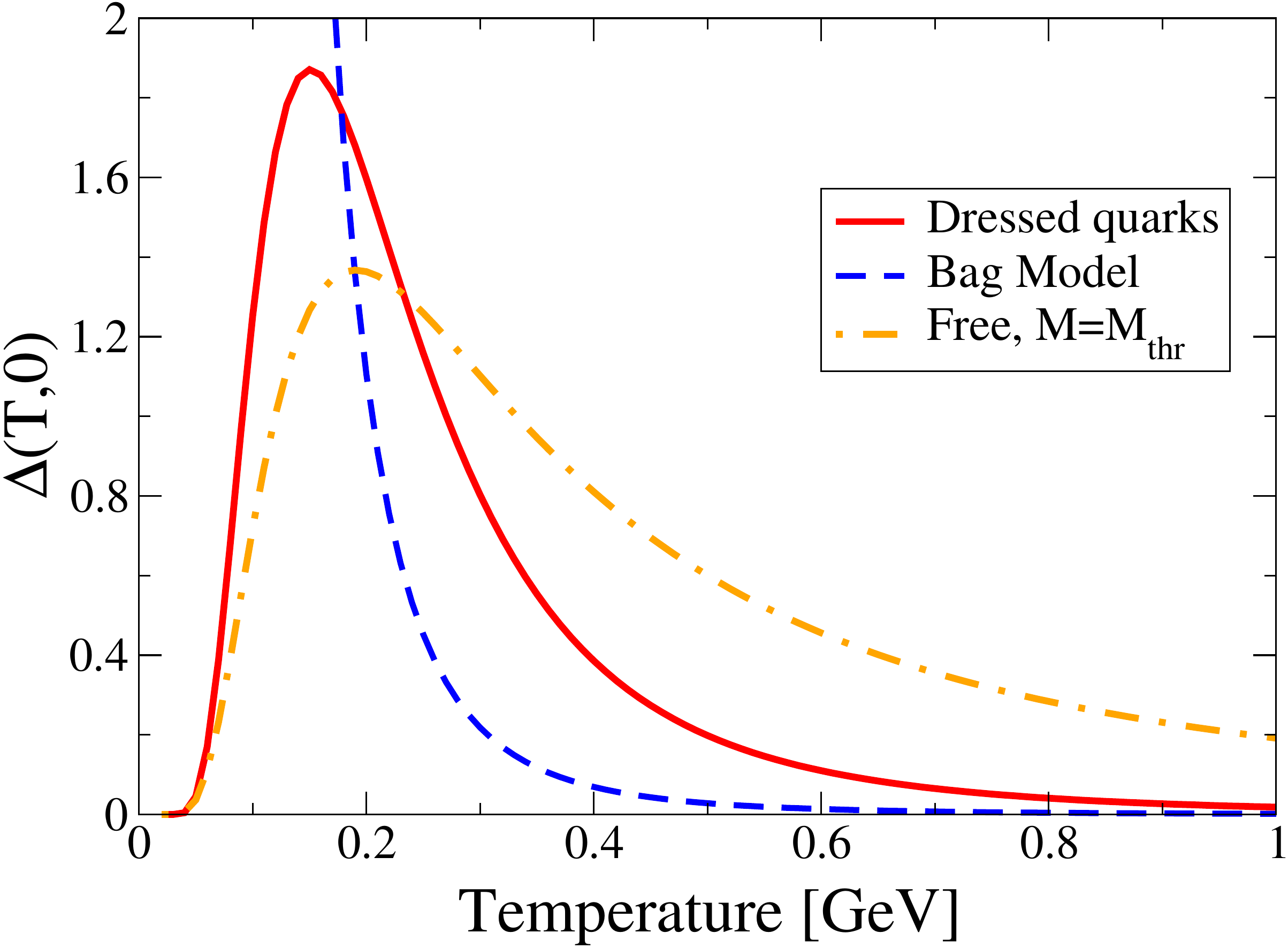}
\caption{\label{fig2}. Trace anomaly at $\mu=0$ for dressed quarks (solid, red line), for massless quarks 
under a bag pressure (dashed, blue line, with $B=(0.145\,{\rm GeV})^4$), and for massive free quarks 
(dashed-dotted line, with mass $M_{thr}=0.467$ GeV).}
\end{minipage} 
\end{figure}

In Figure \ref{fig1}, we show our results for zero chemical potential, with a comparison between our model, a 
massless bag model and a gas of free massive particles with mass $M_{thr}$. Notice that, in the low temperature 
regime, the model result is quite close to that of a free, massive gas of mass $M_{thr}=0.467$ GeV. 
As the temperature rises, the pressure of the model rises above that of the free gas. We interpret this 
as a consequence of the decreasing mass function, Eq. (\ref{eq:mass-function}), as one roughly expects that 
the average momentum of field excitations should increase with temperature. A difference between two kinds 
of modelling should be expected, since our model has positivity-violating quarks as its elementary constituents.

Figure \ref{fig2} displays our results for the normalized trace anomaly at zero chemical potential. Notice 
that the model approaches the massless (conformally invariant) behavior at high temperature faster 
than the massive free gas.
\begin{equation}
 \Delta (T) = \frac{E-3P}{T^4} = T\frac{\partial}{\partial T}\left(\frac{P}{T^4}\right).
\end{equation}

\begin{figure}[h]
\begin{minipage}{17pc}
\includegraphics[width=17pc]{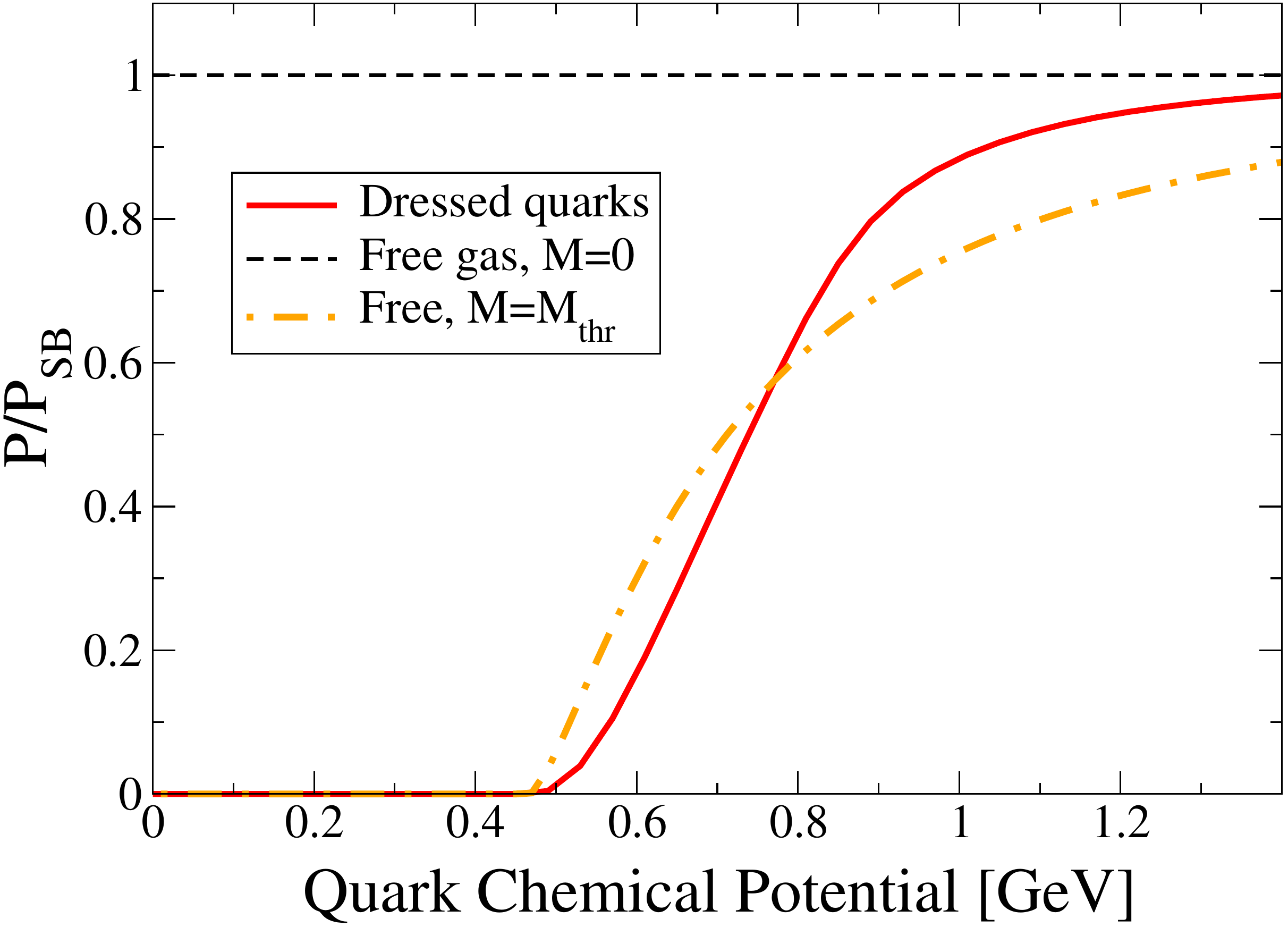}
\caption{\label{fig3} Zero-temperature limit of the pressure as a function of the chemical potential, 
normalized by the Stefan-Boltzmann limit $P_{SB}=N_cN_f\mu^4/(12\pi^2)$. Red, solid line: model result. 
Yellow, dotted-dashed line: free massive gas, with mass $M_{thr}=0.467$ GeV.}
\end{minipage}\hspace{2pc}%
\begin{minipage}{17pc}
\includegraphics[width=17pc]{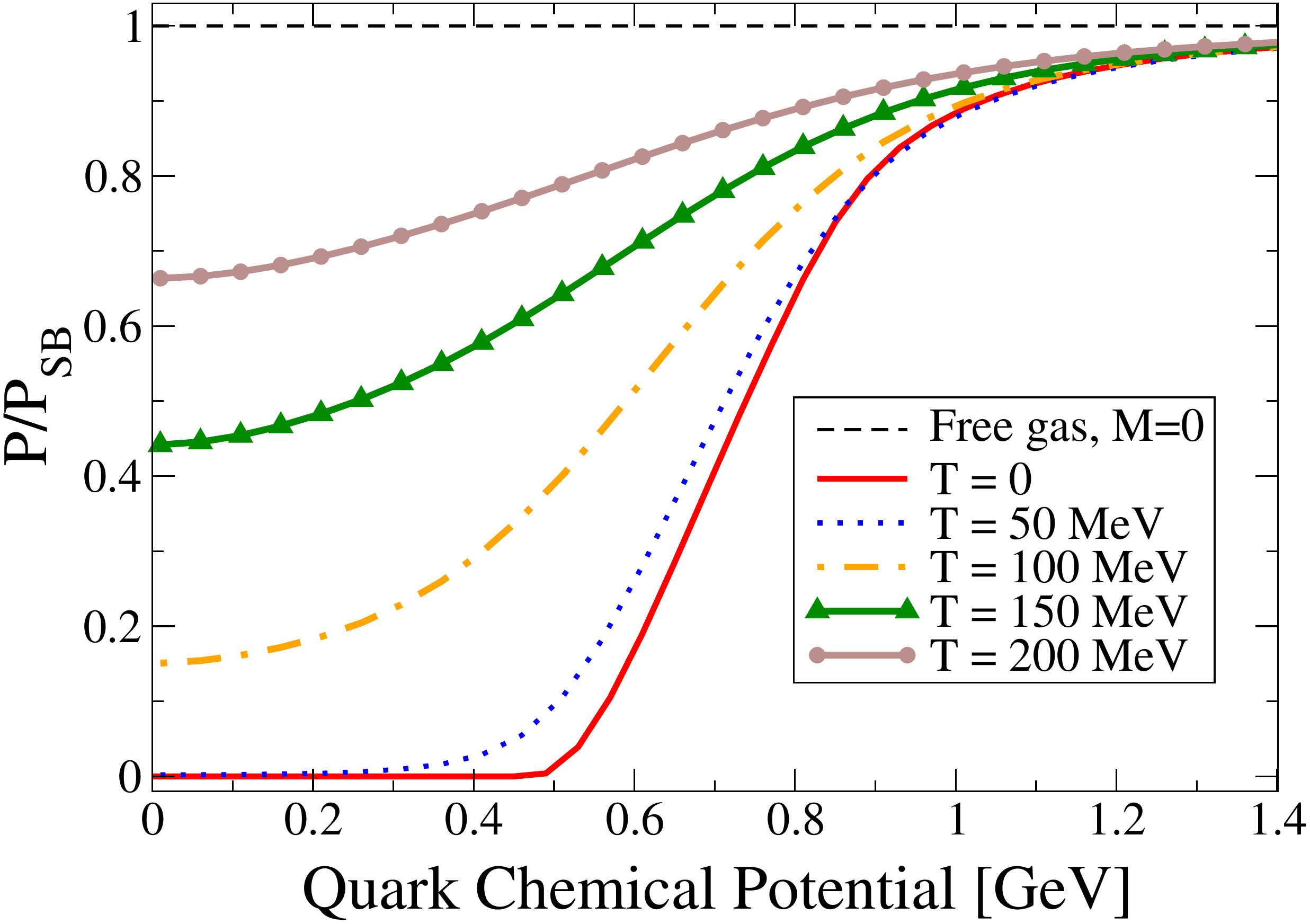}
\caption{\label{fig4} Model pressure as a function of chemical potential for various temperatures. 
Solid, red: $T=0$. Dotted, blue: $T=50$ MeV. Dotted-dashed, yellow: $T=100$ MeV. Solid with triangles, 
green: $T=150$ MeV. Solid with circles, gray: $T=200$ MeV.}
\end{minipage} 
\end{figure}

The zero-temperature, finite chemical potential pressure (\ref{eq:logZ_finite-mu_zeroT}) can be seen in 
Figure \ref{fig3}. Notice that the first excitations appear at a chemical potential 
$\mu\simeq M_{thr}=0.467$ GeV, a mass scale which is not explicitly present in the model action 
(\ref{eq:model}). We thus interpret the threshold mass $M_{thr}$ as a dynamically generated mass scale. 
Notice also that $M_{thr}$ was used for the comparison between the model and a free gas in Figures 
\ref{fig1} and \ref{fig2}, where a low-temperature agreement between the curves appears to be 
satisfactory. This indicates that a description purely in terms of a free gas of mass $M_{thr}$ can only 
be reasonable just above the threshold, but not at higher densities. This is clearly a consequence
of the momentum dependence of the mass function (\ref{eq:mass-function}).

For completeness, we show in Figure \ref{fig4} our results for the pressure at both finite temperature 
and finite chemical potential. As one could expect from general thermodynamical arguments, for a fixed 
value of chemical potential, increasing temperatures lead to higher pressures. Notice that, as soon as 
the temperature is nonzero, thermal activation allows field excitations for arbitrarily low chemical 
potentials. This can be seen from the nonzero values as well as a smoothening of the pressure curve 
already below $\mu=M_{thr}$.

Let us finally notice that all our results are compatible with thermodynamical stability, a feature 
that is not always present in quark models with complex masses (for two studies, see, e.g., 
\cite{Benic:2012ec,Benic:2013eqa}).

\section{Summary}

In this work, we report on a first exploratory investigation of the thermodynamics of the confining quark 
model proposed in \cite{Baulieu:2008fy}. We computed exactly the lowest-order partition function of the 
model at arbitrary temperature and chemical potential. Notice that, although the theory in its lowest 
nontrivial order is quadratic in the fields, it does not correspond to a free theory. Indeed, the quark 
propagator arising from the quadratic theory displays violation of reflection positivity and fits quite 
well lattice results for the quark mass function \cite{Parappilly:2005ei}.

Our results for thermodynamical observables such as pressure and trace anomaly show that the microscopic 
confined degrees of freedom of the model do not correspond to a free gas of massive particles, even at 
lowest order. This can be understood from the dressing of quarks by the underlying gluons, which is 
encoded in the momentum-dependent mass function (\ref{eq:mass-function}). In spite of the effective 
presence of particles with complex masses in the calculation of the partition function, we have not found 
any thermodynamical instabilities in the observables computed.


\subsection*{Acknowledgments}
I wish to thank the organizers of the Hadron Physics Conference in Angra dos Reis (Brazil) for 
having prepared such a nice environment for a scientific meeting. This work was supported by 
the project {\it For Women in Science} of the Fondation L'Or\'eal.


\section*{References}

\end{document}